\documentclass[%
 aip,
 amsmath,amssymb,
 reprint,%
]{revtex4-2}

\usepackage{graphicx}
\usepackage{dcolumn}
\usepackage{bm}

\usepackage[utf8]{inputenc}
\usepackage[T1]{fontenc}
\usepackage{mathptmx}
\usepackage{comment}
\usepackage[dvipsnames]{xcolor}
\usepackage{geometry}
\usepackage{float}

\def\Lg{L_\mathrm{g}} 
\def\cs{c_\mathrm{s}} 

\begin{document}

\preprint{AIP/123-QED}

\title{Simultaneous generation and detection of energetic particle and radiation beams from relativistic plasma mirrors driven at kHz repetition rate}

\newcommand{\LOA}{Laboratoire d'Optique Appliqu\'ee, Institut Polytechnique de Paris, ENSTA-Paris, Ecole Polytechnique, CNRS, 91120 Palaiseau, France}
\newcommand{\Ardop}{Ardop Engineering, Cité de la Photonique, 11 Avenue de la Canteranne, Bât. Pléione, 33600 Pessac, France}
\newcommand{\Weizmann}{Department of Physics of Complex Systems, Weizmann Institute of Science, Rehovot 76100, Israel}

\author{Jaismeen Kaur} \email{jaismeen.kaur@ensta-paris.fr} \affiliation{\LOA} 
\author{Marie Ouill\'e} \affiliation{\LOA} \affiliation{\Ardop}
\author{Dan Levy} \affiliation{\Weizmann}
\author{Louis Daniault} \affiliation{\LOA}
\author{Axel Robbes} \affiliation{\LOA}
\author{Neil Za\"im} \affiliation{\LOA}
\author{Alessandro Flacco} \affiliation{\LOA}
\author{Eyal Kroupp} \affiliation{\Weizmann}
\author{Victor Malka} \affiliation{\Weizmann}
\author{Stefan Haessler} \affiliation{\LOA}
\author{Rodrigo Lopez-Martens} \affiliation{\LOA}

\date{\today}

\begin{abstract}

We report on the first simultaneous measurement of high-order harmonics, relativistic electrons and low divergence proton beams generated from plasma mirrors driven at kHz repetition rate by relativistic-intensity milliJoule-energy femtosecond laser pulses. This setup enables detailed parametric studies of the particle and radiation spatio-spectral beam properties for a wide range of controlled interaction conditions, such as pulse duration and plasma density scale length. This versatile setup should aid in further understanding the collective laser absorption mechanisms at play during the laser-plasma interaction and in optimizing the secondary beam properties for potential applications.
\end{abstract}
\maketitle

\section{\label{introduction}Introduction}

When an intense femtosecond laser pulse is tightly focused on an optically polished solid surface, it generates a thin layer of surface plasma with near-solid density, which becomes highly reflective for the incident laser light, and is therefore often called "plasma mirror" (PM). Progress in high-power laser technology going back almost two decades, in particular the generation of ultra-short laser pulses with high temporal contrast~\cite{DoublePlasmaMirrorDoumy2004, PlasmaMirrorDromey2004, Kalashnikov:05}, has enabled the exposure of PMs to peak electric field amplitudes that can drive relativistic plasma electron motion, making PMs potentially interesting sources of ultrashort energetic particle and radiation beams for applications. For typical lasers emitting around a central wavelength $\lambda_0$ = 800\,nm, this occurs for peak focused intensities $I_0 > 10^{18}\ \textrm{W/cm}^2\ $ or normalized vector potentials $a_0 = \sqrt{I_0\ [\textrm{W/cm}^2]\ \lambda_0^2\ [\mu m^2]/ 1.37\times 10^{18}} > 1$. In this so-called relativistic regime, the highly nonlinear PM response to laser light can lead to generation of high-order laser harmonics (HHG), relativistic electron bunches and energetic ion beams. 

HHG extends into the extreme ultra-violet (XUV) and even up to the X-ray~\cite{RelativisticHHGDromey2006} spectral ranges and corresponds to a train of attosecond pulses in the time domain \cite{RevHHGfromPlasmaMirrors_Thaury2010,Attosecond_Nomura2008,SpatioTemporal_Chopineau2021}. Two main generation mechanisms have been identified in simulations and experiments: coherent wake emission (CWE) \cite{CWE_Quere2006} and relativistic oscillating mirror (ROM) \cite{ROM_Lichters1996, HHGoverDensePlasmas_Baeva2006}. In CWE, the laser field injects vacuum-heated or so-called "Brunel" electrons \cite{BrunelAbsorption_1987} from the surface into the bulk of the plasma, forming a propagating electron density peak which excites plasma oscillations in its wake, finally leading to HHG. CWE is an ubiquitous mechanism in all HHG experiments on PMs \cite{CWE_Quere2006} and is efficient even at intensities as low as $I\sim10^{15}\ \textrm{W/cm}^2$. ROM only comes into play at relativistic intensities, where the PM surface itself begins to oscillate at relativistic velocities, leading to a periodic frequency upshifting of the reflected light, resulting in HHG. In the relativistic regime, it is important to take electron motion into more detailed account, as in the more advanced relativistic electron spring \cite{RES_Gonoskov2018} and coherent synchrotron emission \cite{EnhancedRHHG_Brugge2010,CSE_Mikhailova2012,CSE_Dromey2012, HHGscalingLaw_Edwards2020} models. The relative contributions of relativistic HHG (RHHG) and CWE to the total emission depends mainly on the incident laser intensity \cite{RevHHGfromPlasmaMirrors_Thaury2010, HHGscalingLaw_Edwards2020} and the plasma gradient scale length $\Lg$ \cite{LgEffect_Kahaly2013}. These two HHG families are easily distinguishable experimentally through their different spectral widths, phase properties and divergence. 

\begin{figure}
    \newgeometry{}
    \includegraphics[width = 1\textwidth]{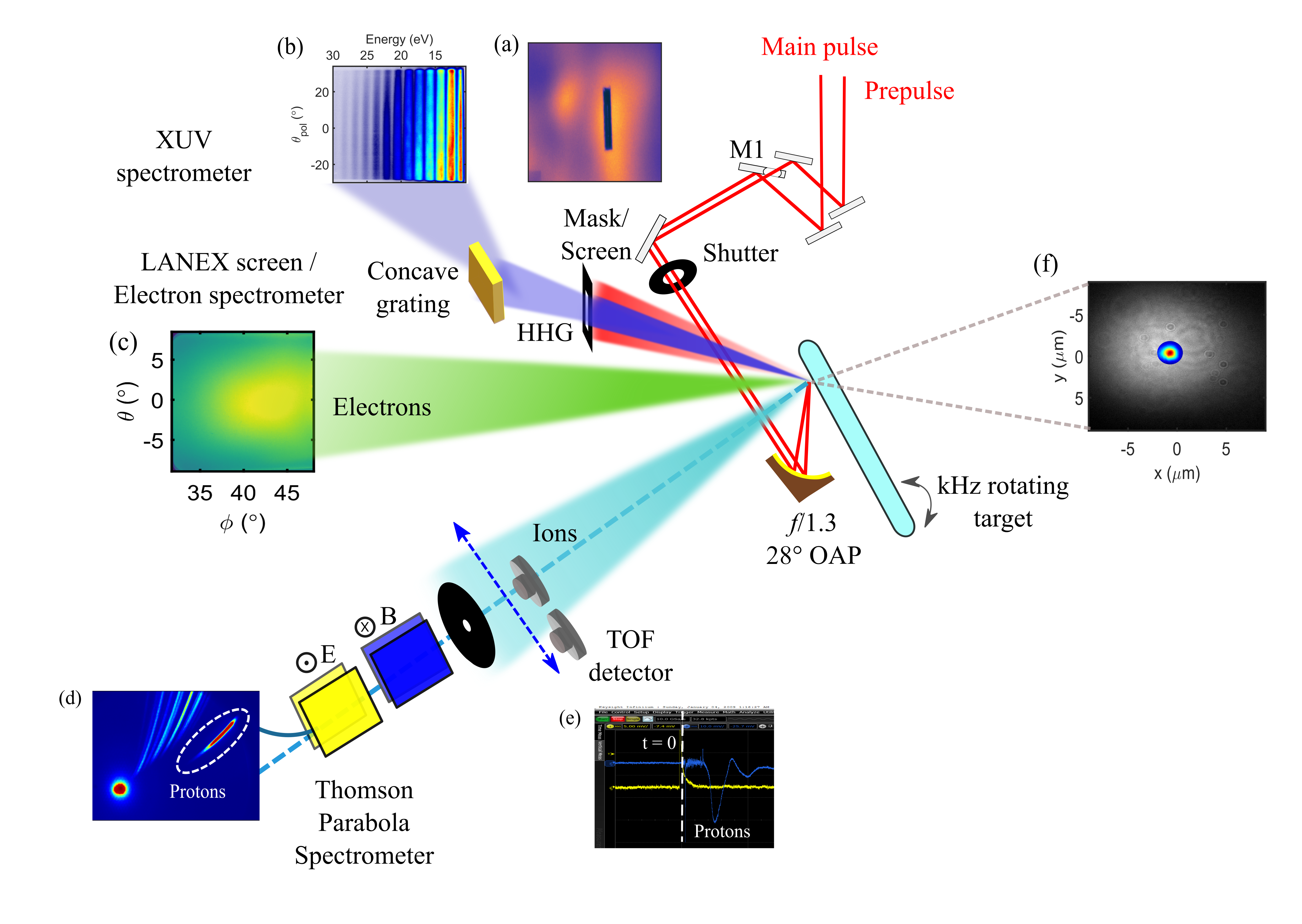}
    \restoregeometry
    \caption{Experimental set-up for simultaneous detection of HHG, electrons and protons emitted from a relativistic PM prepared by a weak Pre-pulse and driven by an intense Main pulse. (a) Reflected laser near-field, (b) HHG spectrograph, (c) spatial electron beam profile, (d) parabolic ion tracks and (e) ion TOF trace are the typically recorded images by the respective diagnostics. (f) On-target Pre-pulse (grey scale) and Main pulse (color scale) spatial beam profiles. M1: Holey mirror}
    \label{fig:expt_setup}
\end{figure}

In addition to HHG, PMs can also emit ultrashort relativistic electron bunches. It is possible to access different acceleration regimes by controlling the plasma conditions. For relatively long density gradients $(\Lg > \lambda_0)$, electrons are accelerated to multi-MeV energies by a laser wake-field produced inside the plasma gradient itself, a process which is only observed with few-optical-cycle driving pulses \cite{LWFAfromPM_Zaim2019}. For sharper density gradients $(\Lg \ll \lambda_0)$, electrons ejected out of the plasma surface co-propagate with the reflected laser field in vacuum and subsequently gain large amounts of energy by vacuum laser acceleration \cite{VLAfromPlasmaMirrors_Thevenet2016}. The energy and charge of the electron beam emitted in this regime have been found to be correlated to RHHG emission \cite{ElectronAccFromOverdensePlasmas_Thevenet2016,IdentCouplingMechnisms_chopineau2019, SHHGelectrons_haessler2022} and anti-correlated to CWE \cite{HHGelectronsAntiCorr_Bocoum2016}. 

Most of the experimental ion acceleration studies from PM have been focused on the beams emitted from the back-side of the target through the well-known target normal sheath acceleration (TNSA) mechanism \cite{ProtonsInSolids_Wilkis2001}. TNSA-based ion sources show large beam divergences, varying from $10^\circ$ to $30^\circ$ depending on the driving laser and target parameters \cite{ProtonsFromPetaWattLaser_Snavely2000, LaserIonAccReview_Macchi2013}. Often, additional collimation devices and special target geometries \cite{TargetEnggforProtons_Kar2008, ProtonsInSolids_Wilkis2001}, along with complicated target fabrication and handling are required to keep the beam collimated. However, in case of front-side ion acceleration, a different acceleration mechanism driven by Brunel electrons has been found to be more efficient than TNSA for the steepest plasma density gradients \cite{FrontSideProtonAcc_Hou2009, BrunelProtonAcc_Veltcheva2012}. Acceleration takes place during the time of interaction of the laser pulse with the plasma and results in extremely low-divergence proton beams \cite{CollimatedMeVProtons-Levy2021}. Such exceptional beam properties, useful for numerous potential applications \cite{LaserAccIonsAppl_Torrisi2006, ProtonsForMedIsotope_Fritzler2003} from a fairly simple target geometry warrants further exploration of this particular acceleration regime. 

Here, we report on the recent progress on HHG, electron and proton acceleration from relativistic PMs driven at kHz repetition rate. At the heart of this work is the simultaneous measurement of these three signals with a controlled surface plasma density gradient. A simultaneous measurement enables us to observe direct correlations between the signals whilst ensuring identical interaction conditions on target. These multiple diagnostics are a fine in-situ probe into the collective plasma dynamics and can be a powerful tool to better understand the rich underlying nonlinear physics. Previously such measurements have been limited to HHG and electrons \cite{HHGelectronsAntiCorr_Bocoum2016, IdentCouplingMechnisms_chopineau2019,  SubLaserCycle_chopineau2022, SHHGelectrons_haessler2022}. We extend the previous work by additionally measuring the accelerated proton beams, characterizing them spectrally as well as spatially. At constant driving pulse energy and focused spatial beam distribution, we can vary the driving pulse duration between 4\,fs and 1.5\,ps, thus scanning over a wide range of intensities $I \sim 10^{16} - 10^{19}\ \textrm{W/cm}^2$ and transitioning from the sub-relativistic to the relativistic regime, while observing all three emission types (HHG, electrons and protons) simultaneously. 

\section{\label{exptSetup}Experimental Setup} 

\subsection{Driving pulses}
The experiments were carried out in the \emph{Salle Noire} laser facility at Laboratoire d’Optique Appliqu\'ee (LOA), delivering waveform-controlled, up to 2.5\,mJ on-target, high-temporal contrast ($> 10^{10}@10\:$ps) laser pulses, centered around $\lambda_0 = 780\:$nm, with tunable pulse duration from 27\,fs to sub-4\,fs \cite{SN2Laser_ouille2019}. While maintaining the same on-target fluence (same energy contained within the same focal volume), the laser pulses can also be stretched up to 1.5\,ps by adding group delay dispersion using an acousto-optic programmable dispersion filter (\textit{Dazzler, Fastlite}) integrated into the laser chain. A schematic layout of the experimental set-up is illustrated in figure \ref{fig:expt_setup}. Under vacuum, the p-polarised Main pulses are focused by an f/1.3 $28^\circ$ off-axis parabolic (OAP) mirror (effective focal length $f=54.4\:$mm) onto a rotating, thick fused optical grade silica target at an incidence angle of $\theta_i = 55^\circ$, to a $\approx 1.8\ \mu$m FWHM spot.

\subsection{Plasma density gradient control}
A spatially overlapped, time-delayed Pre-pulse is created by picking-off $\approx 4\%$ of the Main pulse through a holey mirror, decreasing its size by a factor $\approx2$ in a Galilean telescope, and sending it through a delay stage. It is collinearly recombined with the Main pulse through a second holey mirror (M1 in figure~\ref{fig:expt_setup}) before being focused to a $13\ \mu m$ FWHM spot with the same OAP. Its much larger size compared to the Main pulse leads to a spatially homogeneous plasma expansion on the target surface. The plasma density gradient scale length, $\Lg \approx L_0 + \cs\tau$, is scanned by varying the lead time $\tau$ of the Pre-pulse before the Main pulse. Here, $L_0$ is the scale length increase due to the finite temporal contrast of the Main pulse. The plasma expansion velocity $\cs$ is measured with the spatial domain interferometer (SDI) technique~\cite{SDI_Bocoum2015}, detailed in section~\ref{measurements:SDI}.

The overlapped Pre-pulse (in grey scale) and Main pulse (in color-scale) beam profiles in-focus are shown in figure \ref{fig:expt_setup}(f). Whether at fixed lead time or during a $\tau$-scan, the spatial overlap of these two focal spots is extremely stable thanks to laser's excellent pointing stability and to the use of a retro-reflector and a low-wobble motorized translation stage in the Pre-pulse delay line. The retro-reflector produces an elliptical polarization state of the Pre-pulse, which is remediated by a half-wave plate that rotates the major axis of the ellipse into the p-polarization direction.

\subsection{Solid target and shot sequence}
A rotating target positioner refreshes the target surface at kHz repetition rate with precise control over shooting position and rotation speed, maintaining adequate shot spacing between consecutive laser shots~\cite{SolidTarget_Borot2014}. To ensure identical shot-to-shot interaction conditions, the target surface is aligned with high precision to be perpendicular to the rotation axis with the help of a frequency-stabilized helium-neon laser in a Mach-Zehnder interferometer~\cite{SolidTarget_Borot2014}. The residual target fluctuations in the normal direction are maintained $<z_R/5$ peak-to-valley, where $z_R$ is the Rayleigh range of the driving laser. Angular fluctuations are reduced to $<80\ \mu \textrm{rad}$ peak-to-valley, limited by mechanical stress in the target's rotary bearing.

A slow and a high-speed mechanical shutter (\textit{Uniblitz CS45}, 45\,mm diameter, 14\,ms opening time), block the beams in front of the OAP for longer $\sim1$-s durations between acquisitions and control the shot sequence length, respectively. The target controller, made from an Arduino micro-controller board, \textit{(i)} tracks the angular and radial positions of the laser-target interaction point and which of these positions have already been shot at, \textit{(ii)} sets the target rotation speed according to the desired shot spacing (typically $100\ \mu$m), and \textit{(iii)} triggers the shot sequence as well as the diagnostics detailed in section~\ref{diags}. For an acquisition, the target controller triggers the slow shutter opening and sets the target rotation speed. Once an angular position with a fresh target surface has been reached, the fast shutter opens and 14\,ms later the diagnostics are triggered. After the desired acquisition time (typically 100\,ms, thus averaging over 100 consecutive laser shots), the fast shutter closes, the target rotation slows down and the slow shutter closes. Once a full circle of shots at a given radius is completed, the target is translated laterally by typically $100\ \mu$m to commence a new circular shot pattern.

We exclusively use SiO$_2$ as target material in the form of uncoated 5\,inch diameter substrates, polished to $\lambda/10$ flatness on both faces. With a 100\,$\mu$m shot spacing, these targets can in principle take 1.2\,million consecutive shots per face, corresponding to 20\,min of continuous operation at 1~kHz. In practice, we take $\approx1000$ acquisitions of $\approx100$ shots per day of experiment, so that one target lasts approximately one month.

\subsection{\label{diags}Diagnostics}
This experimental set-up has been designed to simultaneously detect three observables resulting from the laser-PM interaction, namely the emission of HHG radiation, energetic electrons, and proton beams. This is achieved with three dedicated diagnostics, illustrated in figure.~\ref{fig:expt_setup}b,c,d and detailed below. These are complemented by a fourth one consisting in recording the reflected laser beam's near-field beam profile, as illustrated in figure~\ref{fig:expt_setup}a. 

\subsubsection{XUV spectrometer}
A rectangular hole in the screen for the near-field beam profile transmits light within the angular acceptance of the XUV spectrometer that spectrally characterizes the HHG emission.  The spectrometer consists of an aberration-corrected, flat-field concave grating (\textit{Hitachi 001-0639}, 600\,groves/mm, incidence angle $\alpha = 85.3^\circ$, optimized for wavelenegths $\lambda \approx 22-124\:$nm) that images in the horizontal dimension the $\sim1\ \mu m$\,size HHG source spot on the PM surface, and spectrally disperses it onto an image plane. Beam propagation is hardly influenced in the vertical dimension which thus provides angular resolution to the resulting spectrograph. A single-stack micro-channel plate (MCP) coupled with a P46 phosphor screen (\textit{Photonis Scientific, APD 1 PS 97X79/32/25/8 I 40:1 P46}) is placed in the image plane. It is finally imaged using a CCD camera (12-bit \textit{PCO Pixelfly VGA}). Figure \ref{fig:expt_setup}(a) shows a typical recorded XUV spectrograph. The MCP is time-gated for 200\,ns synchronously with the laser pulses to suppress background signal from longer incoherent plasma emission. The P46 phosphor screen has a decay time of $\approx 300$\,ns enabling detection at kHz repetition rate. 

\subsubsection{Electron diagnostics}
The angular electron distribution is measured with a scintillating screen \cite{ElectronSpectrometer_Glinec2006} (\textit{Carestream Lanex fast}), placed $\sim 20\:$cm from the point of interaction, covering a small but sufficient angular range between the target normal and the specular direction, for $30^\circ < \phi < 50^\circ$, where $\phi$ is the angle w.r.t the target normal in the plane of incidence. The front side of the Lanex-screen is covered with a $15\ \mu m$ thick aluminium foil to block visible light and low-energy electrons ($E_k < 150\:$keV). Detection of low-energy electrons is also suppressed by the substrate of the Lanex-screen. The green light emission of the Lanex-screen is imaged by a second CCD camera (12-bit, \textit{Pixelink PL-B957U}) through an interferometric band-pass filter. The electron energy spectrum can be measured for $\phi = 38^\circ - 40^\circ$ by inserting a purpose-built electron spectrometer, comprising of a 0.5-mm pinhole and a pair of neodymium magnets ($B_0\approx 80\:$mT over $\approx20\:$mm). The magnetic field map has been measured with a Hall probe and injected into a numerical electron trajectory calculation to calibrate the electron energy vs. the position along the trace of deflected electrons on the Lanex-screen. 

\subsubsection{Proton diagnostics}
Accelerated protons along the target normal are spectrally characterized using an in-house built Thomson parabola spectrometer (TPS). A $300\ \mu m$ pin-hole placed at about 60\,cm from the point of interaction selects protons emitted along the target normal direction, which are then dispersed in energy and charge-mass ratio by a 2\,kV/cm electric and an anti-parallel 300\,mT magnetic field, created by 50\,mm long electrodes and 15\,mm long magnets, respectively. Figure \ref{fig:expt_setup}(c) shows a typical recorded TPS trace. Protons can be easily distinguished from the heavier ions due to their unique charge-to-mass ratio. The TPS trace is recorded with another MCP-Phosphor screen assembly imaged by a third CCD camera (12-bit, \textit{Pixelink PL-B957U}).

Alternatively, angularly resolved proton spectra can be measured with a movable time-of-flight (TOF) detector. It consists of a 6\,mm diameter MCP, placed $\approx37.5\ \textrm{cm}$ from the point of interaction, and mounted on a linear translation stage moving parallel to the target surface. This TOF-MCP has an angular acceptance of $0.9^\circ$, small enough to neglect the convolution owing to the finite size of the detector. The transient MCP current is read out using a \textit{KEYSIGHT MSOS804A} 8\,GHz oscilloscope capable of individually recording all $\sim100\:$ shots in one acquisition, when operated in segmented memory acquisition mode. Figure \ref{fig:expt_setup}(d) shows a typical recorded TOF trace. 

\section{\label{measurements}Experimental Results}

\subsection{\label{measurements:SDI}Plasma density gradient measurement}

The plasma density profile on the target surface is a critical experimental parameter, be it for HHG~\cite{zepf_role_1998,Roedel2012ultrasteep,LgEffect_Kahaly2013,IdentCouplingMechnisms_chopineau2019,SHHGelectrons_haessler2022}, electron acceleration~\cite{VLAfromPlasmaMirrors_Thevenet2016,IdentCouplingMechnisms_chopineau2019,LWFAfromPM_Zaim2019,SHHGelectrons_haessler2022}, or proton acceleration~\cite{CollimatedMeVProtons-Levy2021}. Its control requires a high temporal contrast of the femtosecond driving laser, typically such that its intensity remains $<10^{10}\:$W/cm$^2$ until $<10\:$ps before the pulse peak. This is 2 orders of magnitude below the intensity required to start field-ionizing\cite{delone_tunneling_1998, bauer_ejection_1997} SiO$_2$, and also excludes possible multi-photon processes. If this is the case, a femtosecond Pre-pulse can create a preplasma with an initially quasi-step-like density profile, with an electron density $n_0$ in the half-space $x<0$, the $x$-direction being perpendicular to the target surface. The free expansion of such a plasma can be described by a 1D isothermal (i.e. constant electron temperature $T_\mathrm{e}$) model~\cite{mora_plasma_2003} yielding a time-dependent exponential electron density profile that starts dropping at $x=-c_\mathrm{s}t$ from $n_0$ towards vacuum like
\begin{equation}
    n(x,t) = n_0 \exp\left[-\frac{x}{\cs t} -1\right], 
\label{eq:densityprofile}
\end{equation}
with a gradient scale length $\Lg=\cs t$ that linearly increases with the so-called ion sound velocity $\cs = \sqrt{Z k_\mathrm{B}T_\mathrm{e}/m_\mathrm{i}}$, where $Z$ is the ion charge state, $k_\mathrm{B}$ the Boltzmann constant, and $m_\mathrm{i}$ the ion mass. This implies that, as long as the absorption coefficient of the Pre-pulse by the target surface is intensity-independent, so that the resulting local electron temperature $T_\mathrm{e}$ is proportional to the local Pre-pulse fluence $F$, the local expansion velocity $\cs$ at every position is proportional to $\sqrt{F}$. 

\begin{figure}
    \includegraphics[width=0.5\textwidth]{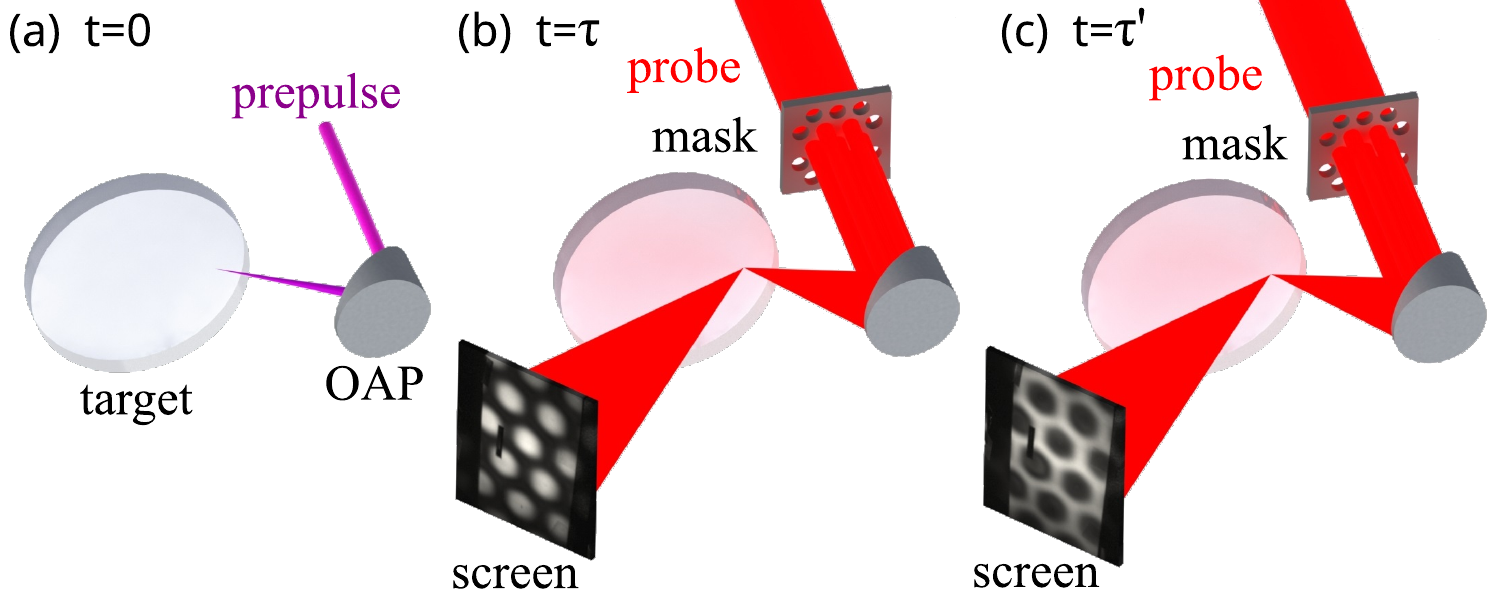}
    \caption{Schematic representation of the experimental SDI measurement. (a) The Pre-pulse initiates the plasma expansion. (b) The probe beam, obtained by shaping the Main pulse with a mask into a hexagonal pattern of smaller beams, is delayed and focused onto the expanding plasma and gets reflected. At certain Pre-pulse lead times $\tau$, the reflected near-field beam profile recreates the pattern imposed by the mask. (c) At certain Pre-pulse lead times, $\tau'$, this pattern is inverted, indicating a phase shift of $\pi$ imposed on the reflected probe beam by the expanding plasma. The images displayed on the "screen" are actual recorded data from a typical SDI measurement.}
    \label{fig:SDIprinciple}
\end{figure}

As a daily routine measurement, this characteristic velocity $c_\mathrm{s}$ of the plasma expansion initiated by the Pre-pulse is determined with the SDI technique introduced by Bocoum \textit{et al.}~\cite{SDI_Bocoum2015}. The principle of SDI is to interfere light reflected off the expanded plasma with light reflected off an unperturbed part of the target surface. The light is considered to reflect on the critical-density surface at depth $x_\mathrm{c}$, where the electron density is $n_\mathrm{c}\cos^2\theta_i$, with $n_\mathrm{c}=\omega_0^2 m_\mathrm{e} \epsilon_0/e^2$, the laser carrier angular frequency $\omega_0$, the electron mass $m_\mathrm{e}$, the vacuum permittivity $\epsilon_0$ and the electron charge $e$. The phase shift due to the reflection at $x = x_\mathrm{c}$ rather than at the initial target-vacuum boundary at $x=0$ is given by (as in Bragg reflection) 
\begin{equation}
    \phi = 2\frac{\omega_0}{c} x_\mathrm{c}\cos\theta_i.
    \label{eq:phiofxc}
\end{equation}
Dephasing due to propagation through the low-density tail of the plasma is at least an order of magnitude smaller and is thus neglected. The measured phase shift as a function of the Pre-pulse lead time $\tau$ thus tracks the position $x_\mathrm{c}(\tau)$ of the critical density surface. Imposing the exponential shape of equation~\ref{eq:densityprofile} (where time $t$ is equivalent to $\tau$) for the plasma density profile, this corresponds to the gradient scale length
\begin{equation}
    \Lg(\tau) = \frac{x_\mathrm{c}(\tau)}{\ln\left[n_0/(n_\mathrm{c}\cos^2\theta_i)\right]-1}.	  
    \label{eq:Lgofxc}
\end{equation}

Experimentally, SDI is implemented  by inserting a periodic transmission mask of period $a$ (in our case a hexagonal pattern of 3\,mm holes with a period $a=4\:$mm) into the near-field of the main beam, as shown in figure~\ref{fig:SDIprinciple}, thus transforming it into the probe beam for SDI. Upon focusing by the OAP, this turns into its far-field diffraction pattern on the target surface. In the paraxial approximation, this is given by the probe beam's spatial Fourier-transform, which is again a hexagonal pattern of "probe spots" (each of the same $\approx 1.8\ \mu$m-FWHM size as the main beam focus) with a period $d=\lambda_0 f/a$, where $\lambda_0$ is the light wavelength. For $\lambda_0=800\:$nm we thus obtain $d=10\:\mu$m. This means that the central (zeroth diffraction order) probe spot overlaps with the center of the Pre-pulse and thus with the nearly homogeneously expanding plasma with which the Main pulse would interact, while the surrounding hexagon of (first diffraction-order) probe spots overlaps with the wings of the 13-$\mu$m Pre-pulse spot, where the Pre-pulse fluence $F_1$ has dropped to $\approx10$\% of its peak value $F_0$.

The expanding plasma thus induces a phase shift $\Delta\phi=\phi_0-\phi_1$ between the central (zeroth order) and the surrounding hexagon of (first order) probe spots. In the simplest case, the first order spots probe an unaffected target surface, so that $\phi_1=0$. 
In our case however, the first-order spots reflect off an expanding plasma "launched" by a finite local Pre-pulse fluence of $F_1\approx0.1F_0$, with the fluence $F_0$ at the Pre-pulse center. Two assumptions are now required: \emph{(i)} Relying on the 1D isothermal expansion model mentioned above, we can write how much smaller the local expansion velocity at the location of the first order probe spots, $\cs^1$, will be compared to its value, $\cs$, at the center: $\cs^1 = \sqrt{F_1/F_0}\cs$. \emph{(ii)} We set the maximum plasma density $n_0$ to be the same everywhere. By combining equations~\ref{eq:phiofxc} and \ref{eq:Lgofxc}, we deduce that
the gradient scale length at the Pre-pulse center, $\Lg$, is proportional to the measurable phase shift $\Delta\phi$:
\begin{multline}
    \Lg(\tau) = \alpha \, \Delta\phi(\tau), \label{eq:Lg} \\
    \mathrm{with}\:\alpha = \frac{\lambda_0/(4\pi\cos\theta_i)}{\left\{\ln\left[n_0/(n_\mathrm{c}\cos^2\theta_i)\right]-1\right\} \left(1-\sqrt{F_1/F_0}\right)}.
\end{multline}

\begin{figure}
    \includegraphics[width=0.5\textwidth]{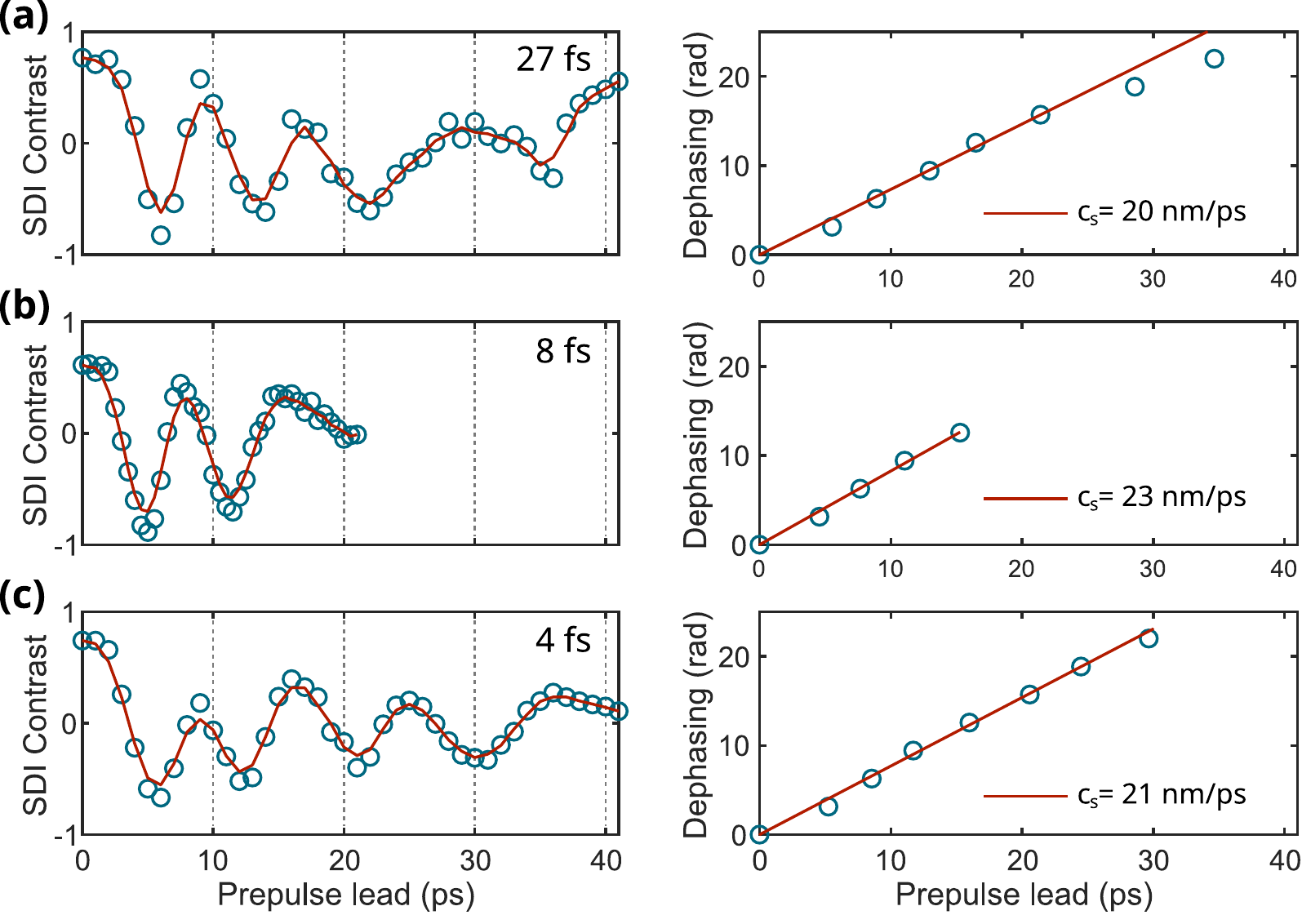}
    \caption{Experimental SDI contrast curves (left, circles are raw data, lines are smoothed data) and extracted dephasing curves curve $\Delta\phi(\tau)$ (right, circles are extracted data, lines are linear fits including points with $\tau<20\,$ps) for SDI scans obtained for driving laser pulse durations of 27\,fs (a), 8\,fs (b) and 4\,fs (c).}
    \label{fig:SDIcompare}
\end{figure}

The measurement of $\Delta\phi$ is made by observing, through an interferometric band-pass filter around $\lambda_0=800\:$nm, the modification of the probe beam's near-field profile (cf. Fig.\ref{fig:expt_setup}a). As demonstrated in ref.~\cite{SDI_Bocoum2015} and illustrated in figure~\ref{fig:SDIprinciple}, the pattern reverses every time the phase shift $\Delta\phi$ reaches an integer multiple of $\pi$. The pattern reversal is easily extracted from the recorded near-field images by tracing the "SDI contrast" defined as $C=[S_\mathrm{b}(\tau)-S_\mathrm{d}(\tau)]/[S_\mathrm{b}(\tau)+S_\mathrm{d}(\tau)]$, where $S_\mathrm{b}$ and $S_\mathrm{d}$ are the integrated signals over selected image areas that are initially (at $\tau=0$) bright and dark, respectively. As visible in the examples shown in the left column of figure~\ref{fig:SDIcompare}, the SDI contrast oscillates and its extrema mark the times, $\tau$, when the dephasing, $\Delta\phi$, has increased by $\pi$. The resulting dephasing curve, $\Delta\phi(\tau)$, is found to be linear (at least for the first $\approx20\:$ps or so), which through equation~\ref{eq:Lg} corresponds to a linearly increasing $\Lg = \cs\tau$, as expected from the 1D isothermal expansion model. Its slope is easily found by a fit on the linear range of data points and yields the sought-after expansion velocity, $\cs$, that is then used to calibrate the gradient scale length in experiments.

Although it only contributes logarithmically in equation~\ref{eq:Lg}, the choice of the $n_0$-value is a significant source of uncertainty. Barrier-suppression intensities for field ionization~\cite{delone_tunneling_1998, bauer_ejection_1997} predict ion charge states for the Pre-pulse and SDI probe pulse intensities in our experiments, which  correspond to $n_0\approx100 n_\mathrm{c}$ only. Field ionization may, however, not be the dominant ionization mechanism. Indeed, once the target is ionized, the laser only interacts inside the plasma skin depth, which is much shorter than a wavelength for a strongly overdense plasma. Then, ionization proceeds through collisional processes in the plasma bulk, possibly further increasing the ion charge states. Estimating the plasma ionization state (which may very well be inhomogeneous and time-dependent) is thus not straightforward. The $\cs$-values shown in figure~\ref{fig:SDIprinciple} are all obtained with $n_0=300 n_\mathrm{c}$, corresponding to fully ionized SiO$_2$.

Figure~\ref{fig:SDIcompare} shows a series of SDI measurements that answers the question whether the plasma expansion initiated by the Pre-pulse varies with laser pulse duration. Since neither the energy nor the spatial profile of the Pre-pulse vary with pulse duration, its \emph{fluence} remains the same ($\approx50\:$~J/cm$^2$) and only its \emph{intensity} varies. The question thus comes down to whether the Pre-pulse absorption coefficient is intensity-dependent, which would lead to varying $T_\mathrm{e}$ and thus varying $\cs$. The telescope and wave plate traversed by the Pre-pulse add a group delay and a third-order dispersion of $\approx260\:$fs$^2$ and $\approx200\:$fs$^3$, respectively. As a consequence, the 27\,fs pulse is stretched to 38\,fs with a peak intensity of $1.1\times10^{15}\:$W/cm$^2$, the 8\,fs pulse is stretched to $\approx100\:$fs with a peak intensity of $0.5\times10^{15}\:$W/cm$^2$, and the 4\,fs pulse is stretched to $\approx200\:$fs, with a profile dominated by a $\approx40\:$fs spike with a peak intensity of $0.7\times10^{15}\:$W/cm$^2$. As visible in figure~\ref{fig:SDIcompare}, the three cases lead to very similar expansion velocities, with a variation clearly below the uncertainty of the SDI method itself. We conclude that the $\cs$-value is independent of the Pre-pulse duration in the range covered by our experiments. This intensity independence corroborates the proportionality $\cs\propto\sqrt{F}$ used in the SDI-analysis. 

The slope of the $\Delta\phi(\tau)$-curves decreases for longer delays, $\tau>20\:$ps, indicating a decreasing velocity, $\cs$, and thus a cooling of the plasma electrons through transfer of energy to the ions. The $\Lg$-values relevant to (most of) our PM experiments are however reached before these effect set in, such that the linear evolution, $\Lg=\cs\tau+L_0$, remains valid.

\subsection{\label{measurements:shortL}Short and intermediate plasma density gradients}

\begin{figure}
    \newgeometry{}
    
    \includegraphics[width=1\textwidth]{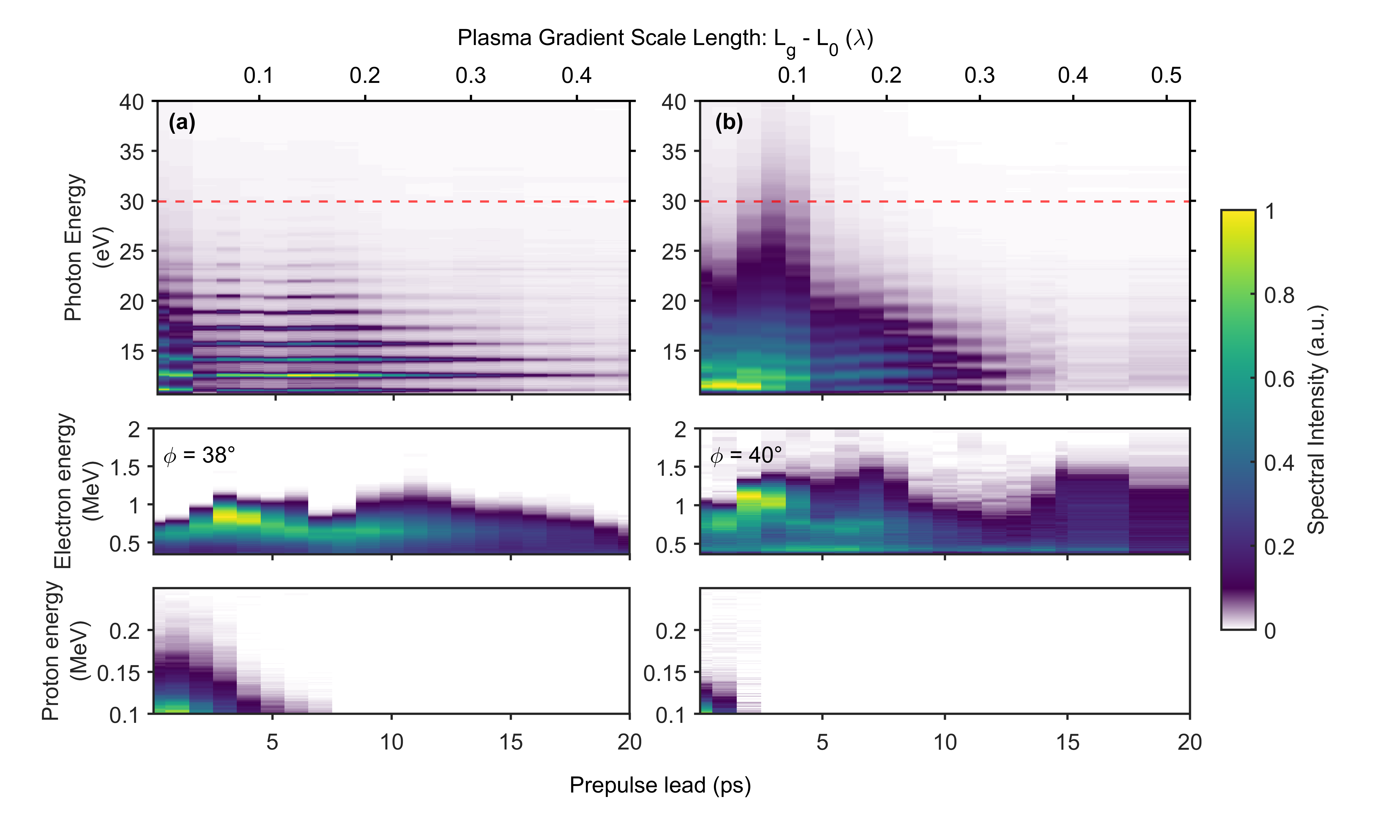}
    \vspace*{-1cm}
    \caption{Simultaneously measured angularly integrated HHG spectrum (top), electron spectrum measured for $\phi  = 38^\circ- 40^\circ$ (middle), and proton spectrum at $\phi = 0^\circ$(bottom) as a function of the plasma density gradient scale length, for different driving pulse durations (a) 29\,fs $(a_0 = 1.0)$, and (b) 4\,fs $(a_0 = 2.1)$. The dotted red line marks the photon cut-off energy for CWE-HHG.}
    \label{fig:delayscans}
    \restoregeometry
\end{figure}

All data reported here are acquired and integrated over 100\,ms long bursts of laser pulses at 1\,kHz repetition rate. Figure \ref{fig:delayscans} shows the simultaneously measured HHG, electron and proton spectra for a range of short to intermediate plasma density gradients, with relativistically intense main driving pulses of (a) 29\,fs ($a_0 = 1.0\ \textrm{for}\ \lambda_0 = 800\ \textrm{nm}$) and (b) 4\,fs ($a_0 = 2.1\ \textrm{for } \lambda_0 = 780\ \textrm{nm}$) duration. The gradient scale length is calculated, up to the small offset $L_0$, as $\Lg-L_0=\cs\tau$ (top axis) from the plasma expansion velocity $\cs = 18\ \textrm{nm/ps}$, measured as described in section~\ref{measurements:SDI}, and the Pre-pulse lead time $\tau$ (bottom axis). Each of the scans in figure \ref{fig:delayscans} have been individually normalized to its maximum value and hence the spectral intensity cannot be directly compared. The HHG spectrum (figure \ref{fig:delayscans} top frame) is angularly integrated over the full detected range of $[-35\ \textrm{mrad}, + 35\ \textrm{mrad}]$. The spectral response of the MCP has not been taken into account so as to enhance the visibility of higher harmonic orders. We can identify the two regimes for HHG: CWE-HHG at steep gradients $(\Lg-L_0 < 0.03\lambda_0)$ and RHHG at more gentle gradients of $\Lg-L_0 \approx 0.03\lambda_0$-- $0.2\lambda_0$ \cite{IdentCouplingMechnisms_chopineau2019, SHHGelectrons_haessler2022}. In the CWE regime, the intensity-dependence of the Brunel electron trajectories leads to a temporal aperiodicity of the emitted attosecond pulse train and, hence, to broadened harmonics in the spectral domain \cite{CWE_Quere2006, IdentCouplingMechnisms_chopineau2019}. In contrast, the RHHG emission in the moderately relativistic regime ($a_0\sim 1$) is Fourier-limited \cite{HHGphaseProperties_Quere2008}, corresponding to narrower spectral widths of the harmonics as compared to the CWE-HHG, as observed in figure \ref{fig:delayscans}(a) (top frame) with 29\,fs driving pulse duration. However, this transition between the CWE and RHHG regimes becomes less distinct with shorter driving pulse durations ($a_0>1$) since the harmonics emerge in the spectrograph from the interference between fewer attosecond pulses. In figure \ref{fig:delayscans}(b), for a 4\,fs driving pulse, around an optimal $\Lg-L_0 \approx 0.08\lambda_0$, the HHG spectrum extends well beyond the CWE cutoff of 30\,eV for an SiO$_2$ target, which is definitive proof of RHHG.

The electron energy spectrum (figure \ref{fig:delayscans}, middle frame) is simultaneously measured at $\phi = 38^\circ$-- $40^\circ$. These positions do not sample the center of the electron beam, which is located a few degrees closer towards the specular direction (see Fig.~\ref{fig:expt_setup}c), but they allow simultaneous detection of the HHG emission. At the steepest gradients, where the CWE-HHG is optimum, low electron energy and charge is measured for both driving pulse durations of 29\,fs and 4\,fs. This anti-correlation between the electron beam energy and charge and CWE-HHG has also been previously observed at sub-relativistic intensities with 30\,fs driving laser pulses~\cite{HHGelectronsAntiCorr_Bocoum2016}. We measure electrons up to 1.5\,MeV in energy in the relativistic regime for $a_0>1$. Higher energies are expected in the center of the electron beam. The apparent dip in the electron spectra at the optimal gradient for RHHG is suspected to be linked to a moving electron beam.

\begin{figure}
    \includegraphics[width=0.5\textwidth]{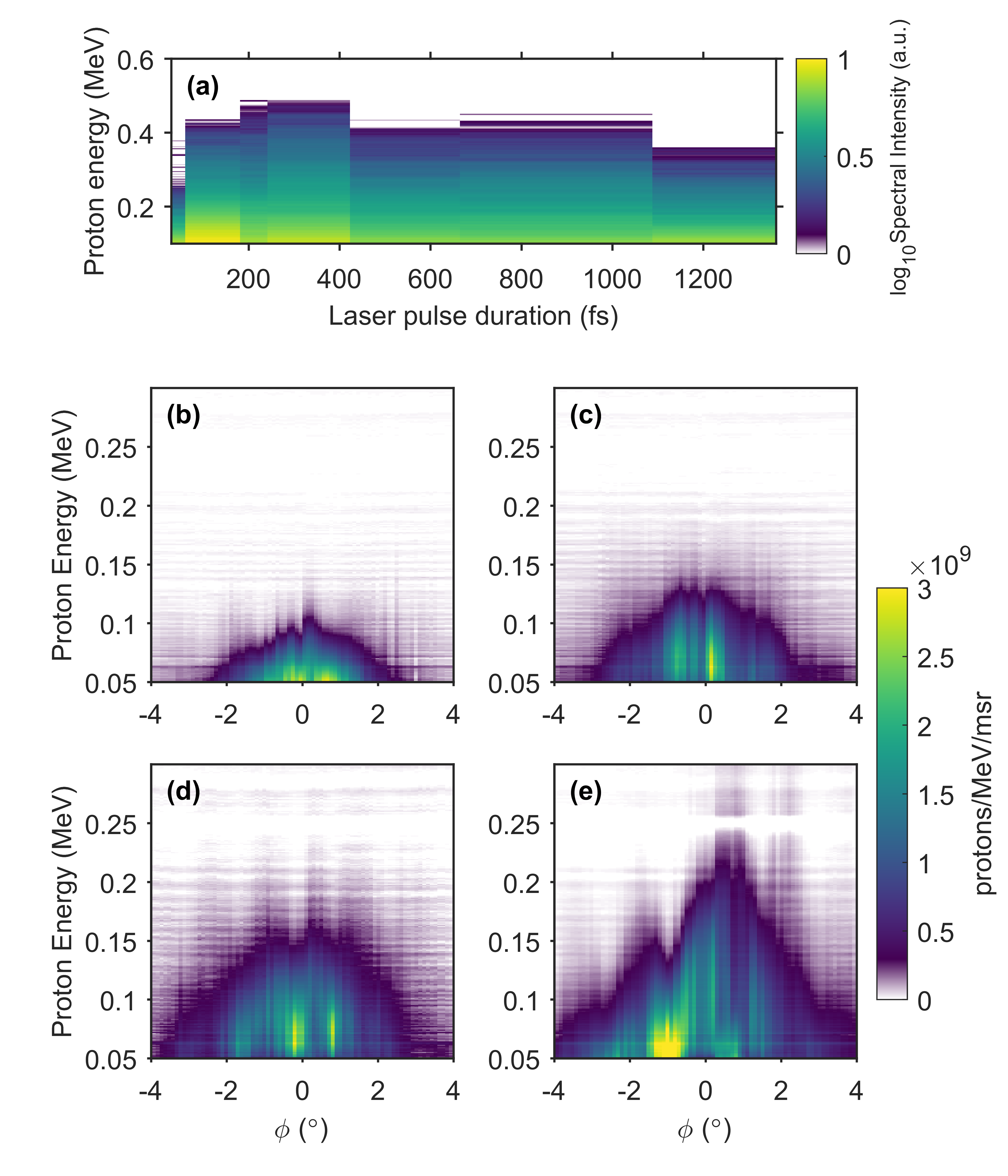}
    \caption{Proton energy spectrum measured with TPS at the steepest density gradients for different driving pulse durations (a). Angular proton energy spectrum obtained for (b) 4\,fs, (c) 7\,fs, and (d) 29\,fs and (e) 200\,fs driving pulses measured by the moving TOF detector. $\phi = 0^\circ$ is the target normal direction.}
    \label{fig:Protons_TPS_TOFscans}
\end{figure}

Figure \ref{fig:delayscans} (bottom frame) shows the proton energy spectrum simultaneously measured with the TPS along the target normal direction ($\phi=0^\circ$). We observe that for both driving pulse durations, the proton energies benefit from the steepest plasma density gradients and are optimised for the same gradients as CWE-HHG, suggesting that Brunel electrons play an important role in the acceleration mechanism~\cite{BrunelProtonAcc_Veltcheva2012,CollimatedMeVProtons-Levy2021}. If an optimal plasma gradient exists for efficient acceleration of protons, it is below our resolution limit of $L_0 \approx 0.01\lambda_0$. The highest proton energies of 0.25\,MeV are obtained with 29\,fs driving pulses for the steepest density gradients and drop below the detection threshold of 0.1\,MeV for $\Lg>0.1\lambda_0$. The cut-off energy decreases to $\approx0.15\,\textrm{MeV}$ for shorter 4\,fs driving pulses.

Figure \ref{fig:Protons_TPS_TOFscans}(a) shows the proton energy spectrum measured with the TPS for the steepest plasma density gradient and longer driving pulses obtained by adding positive group delay dispersion. We observe an optimum between 100\,fs and 300\,fs, yielding $\approx 0.45\ \textrm{MeV}$ protons. In this acceleration regime, protons are accelerated only during the the interaction time of the pulse, hence they benefit from a longer driving pulse duration, despite the corresponding drop in laser intensity \cite{BrunelProtonAcc_Veltcheva2012,CollimatedMeVProtons-Levy2021}. Figure \ref{fig:Protons_TPS_TOFscans}(b)-(e) shows the angle-resolved proton energy spectrum measured with the moving TOF detector for four different driving pulse durations, 4\,fs (b), 7\,fs (c), 29\,fs (d) and chirped 200\,fs (e), at the steepest plasma gradient. In all cases, we observe a very low $\le 4^\circ$ FWHM divergence of the proton beam. This is significantly lower than the $16^\circ$ FWHM reported earlier \cite{FrontSideProtonAcc_Hou2009}, which may be the result of a higher temporal contrast ratio of the driving laser in our case.

\subsection{\label{measurements:longL}Long density gradients regime}

\begin{figure}
    \centering
    \hspace*{-0.5cm}
    \includegraphics[width=0.55\textwidth]{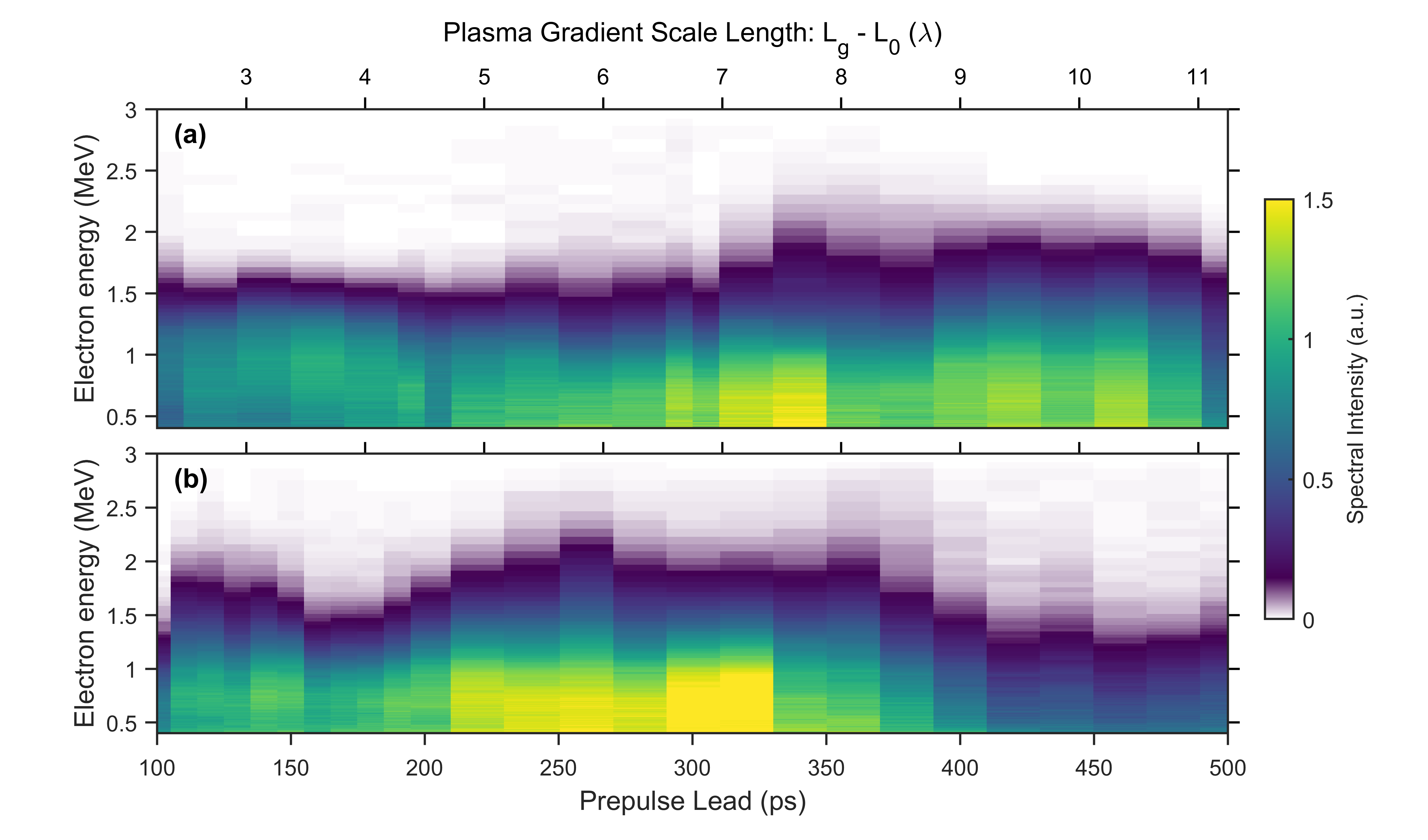}
    \caption{Electron spectrum measured at $\phi  = 40^\circ$ for (a) 7\,fs and (b) 4\,fs driving laser pulse duration and varying plasma density scale length.}
    \label{fig:LWFAelectrons}
\end{figure}

For long density gradients $(\Lg \ge \lambda_0)$, RHHG disappears due to the onset of chaotic electron dynamics in the near-critical-density layer \cite{IdentCouplingMechnisms_chopineau2019} of the plasma density gradient. However, before getting reflected, the laser propagates through the sub-critical tail of the plasma density profile. In the zone of density $\sim n_\mathrm{c}/10$, few-cycle laser pulses $< 10\ \textrm{fs}$ are able to resonantly excite a strong plasma wakefield. Electrons can be injected into this laser wakefield by ionization and then get accelerated to relativistic energies. The plasma wavefront is rotated by the density gradient and electrons are pushed away from the specular direction~\cite{LWFAfromPM_Zaim2019}. Figure \ref{fig:LWFAelectrons} shows the electron spectra measured in this regime for gradient scale lengths up to $\approx 10\lambda_0$, reaching up to 2.6\,MeV. Note that the gradient scale lengths given by the top axis result from the same 18\,nm/ps expansion velocity, as determined for the initial expansion phase with Pre-pulse lead times $<20\:$ps. As discussed in connection with figure~\ref{fig:SDIcompare}, we know that the expansion slows down considerably at longer times, so that the true gradient scale lengths are certainly shorter than those indicated in Figure~\ref{fig:LWFAelectrons}.

\section{\label{summary}Summary}
We have reported the first simultaneous measurements of HHG, electron beams and proton beams from a laser-driven PM. The kHz laser repetition rate allowed exploring a wide range of laser and plasma parameters. We have presented spectral measurements of HHG, electrons and protons as a function of controlled plasma density scale lengths ranging from $0.01\lambda_0 - 0.2\lambda_0$ for three different driving pulse durations: 29\,fs $(a_0=1.0)$, 7\,fs $(a_0=2.0)$, and 4\,fs $(a_0=2.1)$. Such simultaneous measurements vastly expand the level of characterization of relativistic PMs. While ensuring identical interaction conditions on target, we observe direct correlations between the CWE-HHG and Brunel-dominated accelerated proton beams, and the anti-correlated emission of relativistic electron beams. We also observe direct correlation between the RHHG and the emitted electron beams. At extremely long density gradients, only with few-optical-cycle driving laser pulses, we measure up to 2.5\,MeV electrons accelerated by the laser wake-field formed in the gradient. 

We also measured proton spectra for a wide range of driving pulse durations, from 4\,fs up to 1.5\,ps, at the steepest plasma density gradient, and find an optimum pulse duration range for proton yield, between 100\,fs and 400\,fs, under such conditions. These novel measurements of low-divergence $(< 4^\circ\ \textrm{FWHM})$ font-side-accelerated proton beams with a controlled density gradient hold intrinsic value of their own. The exceptional beam properties, along with the possibility of scaling with the driving laser energy to multi-MeV proton energy levels~\cite{CollimatedMeVProtons-Levy2021}, are promising for applications in the field of medicine, for example to produce radio-isotopes~\cite{ProtonsForMedIsotope_Fritzler2003}. 

To conclude, this work is a decisive step towards a highly detailed characterization of fundamental PM dynamics and the optimal generation of energetic particle and radiation beams exploiting the high repetition rate of our laser system.

Additionally, we have also implemented carrier-envelope phase control, with feedback at the full repetition rate of the laser system. This should enable the generation of isolated intense attosecond pulses \cite{IAPfromPM_Boehle2020} with high conversion efficiencies \cite{IIAP_Jahn2019}, by controlling the collective plasma dynamics at a the sub-light-cycle level.  

\begin{acknowledgments}
This work has been supported by the Agence Nationale pour la Recherche (ANR-14-CE32-0011-03 APERO), the European Research Council (ERC ExCoMet 694596) and the European Union's Horizon 2020 research and innovation program (LASERLAB-Europe under grant agreements 654148 and 871124). 
\end{acknowledgments}

\section*{Data Availability Statement}
The data that support the findings of this study are available from the corresponding author upon reasonable request.


\end{document}